\documentstyle[preprint,tighten,eqsecnum,aps]{revtex}

\begin{document}
\draft

\newcommand{\BQ}{\begin{equation}}
\newcommand{\EQ}{\end{equation}}
\newcommand{\BQA}{\begin{eqnarray}}
\newcommand{\EQA}{\end{eqnarray}}
\newcommand{\half}{\frac{1}{2}}
\newcommand{\NN}{\nonumber \\}
\newcommand{\E}{{\rm e}}
\newcommand{\Gmu}{\gamma^{\mu}}
\newcommand{\Gnu}{\gamma^{\nu}}
\newcommand{\gmu}{\gamma_{\mu}}
\newcommand{\gnu}{\gamma_{\nu}}
\newcommand{\gfive}{\gamma_5}
\newcommand{\del}{\partial}
\newcommand{\k}{\mbox{\boldmath $k$}}
\newcommand{\itl}{\mbox{\boldmath $l$}}
\newcommand{\itP}{\mbox{\boldmath $P$}}
\newcommand{\q}{\mbox{\boldmath $q$}}
\newcommand{\p}{\mbox{\boldmath $p$}}
\newcommand{\x}{\mbox{\boldmath $x$}}
\newcommand{\y}{\mbox{\boldmath $y$}}
\newcommand{\Z}{{\bf Z}}
\newcommand{\R}{{\rm R}}
\newcommand{\M}{{\cal M}}
\newcommand{\T}{{\cal T}}
\newcommand{\U}{{\cal U}}
\newcommand{\SS}{{\cal S}}
\newcommand{\PP}{{\cal P}}
\newcommand{\tr}{{\rm tr}}
\newcommand{\Path}{{\rm P}\,}
\newcommand{\dagg}{\mbox{\scriptsize{\dag}}}
\newcommand{\dagpsi}{\psi^{\mbox{\scriptsize{\dag}}}}
\newcommand{\ket}[1]{\left.\left\vert #1 \right. \right\rangle}
\newcommand{\bra}[1]{\left\langle\left. #1 \right\vert\right.}
\newcommand{\ketrm}[1]{\vert {\rm #1} \rangle}  
\newcommand{\brarm}[1]{\langle {\rm #1} \vert}  
\newlength{\bredde}
\def\slash#1{\settowidth{\bredde}{$#1$}\ifmmode\,\raisebox{.15ex}{/}
\hspace*{-\bredde} #1\else$\,\raisebox{.15ex}{/}\hspace*{-\bredde} #1$\fi}
\renewcommand{\thefootnote}{\fnsymbol{footnote}}


%
%
\thispagestyle{empty} 
\begin{titlepage}    
\topskip 4cm
\setcounter{page}{1}
\begin{center}
  {\Large{\bf 
  Quantum description for a chiral condensate disoriented in 
      a certain direction in isospace } }
\end{center}                                   
           
\vspace{0.8cm}
              
\begin{center}
S. Maedan            
   \footnote{ Email address: maedan@tokyo-ct.ac.jp}   
           \\
\vspace{0.8cm} 
{\sl  Department of Physics, Tokyo National College of Technology,
      Tokyo 193-0997, Japan
                                   }
\end{center}                                               
            
\vspace{0.5cm}
              
\begin{abstract}  
\noindent       
We derive a quantum state of the disoriented chiral condensate dynamically,
   considering small quantum fluctuations around a classical chiral condensate
   disoriented in a certain direction  $  \vec n $ in isospace.
The obtained nonisosinglet quantum state has the characteristic features;
   (i) it has the form of the squeezed state,
   (ii) the state contains not only the component of pion quanta in the direction
   $  \vec n $ but also  the component in the perpendicular direction to $  \vec n $
   and (iii) the low momentum pions in the state
   violate the isospin symmetry.
With the quantum state, we calculate the probability of the neutral fraction
   depending on the time and 
   the pion's momentum, and find that the probability has an unfamiliar form.
For the low momentum pions, the parametric resonance mechanism works with the
    result that the probability of the neutral fraction becomes the well known form 
   approximately and that the charge fluctuation is small.
\end{abstract} 
\vskip 1.5cm
\begin{center}
PACS number(s):  11.30.Rd, 12.39.Fe, 25.75.-q, 11.10.Ef
\end{center}
\vfill            
\end{titlepage}
%
%
%
%
%
%
\section{Introduction}

The possibility of realizing the disoriented chiral condensate ( DCC ) has been
   discussed   \cite{rf:AnsRys,rf:BlaKrz,rf:Bjo,rf:KowTay,rf:RajWil395}.
It is expected that the formation of the DCC would be discovered in the experiments
   at Relativistic Heavy Ion Collider ( RHIC ), where an extremely high temperature is
   realized. 
The experimental evidence of the DCC supports the idea of the dynamical
   chiral phase transition in QCD
   which is one of the most important concept in hadron physics.
In the ordinary QCD vacuum, the chiral condensate has isospin $ I=0 $.
However, under the extreme high temperature such as in the experiments at RHIC, 
   it will be possible to form the domain of the DCC.
In the DCC, a chiral condensate does not align in the
   $ < {\bar \psi} \psi > = < \sigma > $ direction in  chiral space.
The chiral condensate is disoriented in a certain direction $\vec n$ in
   isospace,  $ {\vec \pi} = ( \pi_1, \pi_2, \pi_3 ) $.
 
It was pointed out by Rajagopal and Wilczek \cite{rf:RajWil577} for the first time
   that there is  a possibility to form a domain of the DCC.
They carried out numerical simulation in the classical $ O(4) $ linear sigma model
    which is an effective theory of QCD.
Their simulation shows that, after quenching, long wavelength modes of the pion
   fields are amplified.
After quenching, the field $ \phi  = (  \sigma , {\vec \pi} ) $ roll down randomly in
   chiral space from the top of the Mexican hat potential, therefore $\phi$ does not
  align in the $\sigma$ direction in general.
We assume that the DCC, namely, a field configuration  $\phi$ which does not align
   in the  $\sigma$  direction, is formed.
The DCC then relaxes to the true vacuum  $ ( < \sigma > , ~{\vec 0}~ ) $ due to the
   existence of the current quark mass.
That decay of the DCC proceeds with the emission of pions.

Kowalski and Taylor  have studied the quantum state of the DCC \cite{rf:KowTay}.
Since the classical wave picture corresponds to the coherent state, they consider
   the quantum coherent  state of pions.
The coherent state will contain the state of arbitrarily large charge, so they
   suggested that the quantum  state of the DCC has isospin $I=0$.
That isosinglet quantum state leads to the probability of the  neutral fraction
\begin{equation}
  P(f) =\frac{1}{2 \sqrt{f} } ~,
  \label{aa}
\end{equation}
where $ f \equiv N_{\pi^0} /( N_{\pi^0} +N_{\pi^+}+N_{\pi^-} ) $.
On the other hand, in order to describe the DCC, Amado and Kogan \cite{rf:AmaKog}
   have considered  the quantum squeezed state.
They have proposed various squeezed state descriptions of the DCC and studied
   their features. 
Unfortunately, however, the concrete dynamical mechanism of how such coherent
   or (nonisosinglet) squeezed state
   emerges was not discussed\footnote{
   In Ref.\cite{rf:AmaKog}, an isosinglet 
   squeezed state of the pions are derived dynamically by quantizing in the 
   Hartree approximation; $ \phi^2 \rightarrow \langle \phi^2 \rangle (t) $.
   However, a nonisosinglet squeezed state with arbitrary isospin direction is
   given by an ansatz.
    }
   in Ref.~\cite{rf:KowTay} or Ref.\cite{rf:AmaKog}.
So far we have discussed the DCC not aligned in the $\sigma$ direction,  which will be
   a natural consequence of the quench scenario.
One can consider another situation, where a classical chiral condensate aligns in the
    $\sigma$ direction  and oscillates.
In this case, it is shown that the parametric resonance mechanism works 
    \cite{rf:MroMul,rf:HirMinC61,rf:Kai,rf:Ish,rf:DumSca} and an isosinglet
   squeezed quantum state is derived dynamically \cite{rf:HirMinC61}.
The classical chiral condensate has isospin $I=0$ because one has assumed that 
   it aligns in  the  $\sigma$ direction.

By observing the emitted pions from the DCC, one can obtain information on it.
Since what we observe in the experiment is quanta of pions, it is very important
   to study quantum state of the DCC.
Furthermore, the quantum state should be derived dynamically, not by an ansatz.
Such derivation allows us to investigate the property of the DCC quantum state
   and to calculate observable physical quantities.
In this paper, we discuss how the quantum state of the DCC emerges dynamically
   by use of the $O(4)$  linear sigma model.
The formation of a classical DCC, i.e., a classical chiral condensate disoriented in
   a certain direction $\vec n$ in isospace is assumed.
The decay process of the DCC is concerned.
We also study the property of the obtained DCC quantum state.
In order to derive the explicit form of the quantum state, some approximations have
   been made by which
   we obtain a nonisosinglet squeezed quantum state of the DCC.
We have assumed that the classical chiral condensate disoriented in a certain
   direction $\vec n$ in isospace is formed.
According to this assumption, it seems that the quantum DCC state will be described
   by the component of pion quanta in the direction $\vec n$ in isospace.
If so, when $ {\vec n} = ( 1,0,0 ) $, the quantum state is described only by a quantum
    $\pi_1$.
This is, however, a wrong guess.
As shown later, the quantum state contains  not only the component of pion quanta
    in the direction
    $  \vec n $ but also  the component in the perpendicular direction to $  \vec n $.
Obtained quantum state is nonisosinglet, hence there exists the fluctuation of
   the charge, which we estimate.
The probability of the neutral fraction  $ P(f( \k, t ) ) $ depending on the
   momentum $ \k $ of
   pion and the time $t$ is also calculated.
We show that a parametric resonance mechanism \cite{rf:Mae} works for low
    momentum pions $ 0 < \vert \k \vert << m_\pi $.
It is interesting to examine the influence of that mechanism on the probability 
    $ P(f( \k , t ) ) $.

The plan of this paper is the following.
In Sec.~II, the classical theory of the $O(4)$ linear sigma model is presented.
We first assume that a classical field configuration not aligned in the $\sigma$ 
   direction is formed.
The equations of motion for small fluctuations around that configuration are derived.
Although these ( coupled ) equations of motion are complicated, one can rewrite
   them very simple form  by changing the coordinate in isospace.
In Sec.~III, the pions' fluctuations 
   discussed in Sec.~II are quantized.
Neglecting the higher order terms of the quantum fluctuations, we obtain the
   eigenstate of the Hamiltonian,
   which we consider the quantum state of the DCC.
In Sec.~IV, by use of the obtained nonisosinglet quantum state of the DCC, the
   fluctuation of
   the charge $Q$ and the probability  of the neutral fraction  $ P(f( \k, t ) ) $ 
  depending on $ \k $ and $t$ are calculated.
In Sec.~V, we show that the parametric resonance mechanism works for low
   momentum pions, and examine
   the effect of that mechanism on the fluctuation of $Q$ or the probability
   $ P(f( \k, t ) ) $.
Sec.~VI is devoted to the summary and the discussion.
The effects of the quantum state of $\sigma$ or the quantum back reaction are
    discussed.
%
%
%
%
\section{Diagonalization of the equations of motion in isospace}
The lagrangian of the $O(4)$ linear sigma model is
\begin{equation}
 {\cal L}= {1\over2} \partial^\mu \phi \partial_\mu \phi -{1\over4} \lambda
  (\phi \phi -v^2)^2 + H \sigma ,
  \label{ba}
\end{equation}
with $\phi = ( \sigma \, , \vec \pi \,)$.
Because of the explicit chiral symmetry breaking term $H \sigma$ which originates
   from the  current quark mass, the true vacuum is $\phi= ( f_\pi \, , \vec 0 \,)$
  where $f_\pi$ satisfies
   $  \lambda f_\pi ( f_\pi^2 - v^2 ) -H=0 \, $. 
Around the true vacuum, the pions $ {\vec \pi }= ( \pi_1, \pi_2, \pi_3 ) $ have mass 
   $ m_\pi = \sqrt{H / f_\pi}$ and the sigma $\sigma$ has
   $ m_\sigma = \sqrt{ \lambda ( 3 f_\pi^2 - v^2 ) }$.
The charged pions $\pi^{\pm}$ are represented with
   $ ( \mp  \pi_1 + i \pi_2 )/ \sqrt{2} $ and
   the neutral pion $\pi^0$ with $\pi_3$.
Let us consider the field $\phi = ( \sigma \, , \vec \pi \,)$ around the true vacuum
   $ ( f_\pi \, , \vec 0 \,)$,
\begin{eqnarray}
  \sigma (x) &=& f_\pi + \chi (x) ,   \nonumber  \\
  {\vec \pi}(x) &=&  {\vec \pi}(x) .
  \label{bb}
\end{eqnarray}
The equations of motion for $ {\vec \pi} (x) $ and $\chi(x)$ are
\begin{eqnarray}
  &&  \left[  \partial_\mu \partial^\mu + m_\pi^2 + \lambda \{ 2 f_\pi \chi + \chi^2 
   + ( ~ {\vec \pi} \cdot ~ {\vec \pi} ) \} \right] ~ {\vec \pi} = 0 ~~,      \nonumber  \\
  &&  \left[  \partial_\mu \partial^\mu + m_\sigma^2 + \lambda \{ 3 f_\pi \chi + \chi^2 
   + ( ~ {\vec \pi} \cdot ~ {\vec \pi} ) \} \right] ~ \chi 
                      = - \lambda f_\pi  ( ~ {\vec \pi} \cdot ~ {\vec \pi} ) ~. 
  \label{bc}
\end{eqnarray}
To a zeroth approximation, the equations of motion become
\begin{equation}
 \begin{array}{l}
   \left[  \partial_\mu \partial^\mu + m_\pi^2 \right] ~ {\vec \pi}^{(0)} = 0 ,   \\
   \left[  \partial_\mu \partial^\mu + m_\sigma^2 \right] ~ \chi^{(0)} = 0 ,
 \end{array}
    \label{bd}
\end{equation}
where we have neglected the quadratic or cubic terms in the fields.
Since we are interested in the DCC not aligned in the $\sigma$ direction, we assume
   the following  field configuration,
\begin{eqnarray}
   {\vec \pi}^{(0)}(t)  &=&  \pi^{(0)}(t) ~{\vec n} ~,   \nonumber  \\
   \chi^{(0)}  &=&  0 ~,
  \label{bea}
\end{eqnarray}
where 
\begin{equation}
   {\vec n}= ( \sin \theta \cos \phi ,  \sin \theta \sin \phi, \cos \theta ~),
  \label{beb}
\end{equation}
is a fixed unit vector in isospace.
The field ${\vec \pi}^{(0)}$ is assumed to be homogeneous for simplicity.
Within a given DCC domain, this supposition seems reasonable.
Of course, for a DCC domain of finite size in  3 dimensional space, the field
   ${\vec \pi}^{(0)}$  depends on $\x $.
We, however,  consider the homogeneous ${\vec \pi}^{(0)}$ for simplicity.

Next, around these ${\vec \pi}^{(0)}$ and $ \chi^{(0)}$, we consider fluctuations 
   $ {\vec \pi}^{(1)}$ and $ \chi^{(1)} $ which will be quantized in the next section.
Keeping the terms no more than quadratic in the $ {\vec \pi}^{(1)} $ or
   $ \chi^{(1)} $ field \cite{rf:MroMul}, we get the equations of motion describing
   the time development of
   $ {\vec \pi^{(1)} } = ( \pi^{(1)}_1 ,   \pi^{(1)}_2 ,  \pi^{(1)}_3  ) $ and $ \chi^{(1)} $,
\begin{eqnarray}
  && \left[ \partial^2 + m_\pi^2  + \lambda ( {\vec \pi}^{(0)}(t) 
                               \cdot {\vec \pi}^{(0)}(t) ) \right]  {\vec \pi}^{(1)}(x) 
     +  2 \lambda \{ {\vec \pi}^{(1)}(x) \cdot {\vec \pi}^{(0)}(t) \} ~  {\vec \pi}^{(0)}(t)
                                                                                                     \nonumber  \\
      && \hskip3cm  = -2 \lambda  f_\pi ~\chi^{(1)}(x)  ~{\vec \pi}^{(0)}(t) , 
                                                                                           \label{bf}               \\
  && \left[ \partial^2 + m_\sigma^2
             + \lambda ( {\vec \pi}^{(0)}(t) \cdot {\vec \pi}^{(0)}(t) ) \right]  \chi^{(1)}(x) 
                                                  \nonumber  \\
     && \hskip3cm   = - \lambda  f_\pi ~ \{  ( {\vec \pi}^{(0)}(t) \cdot {\vec \pi}^{(0)}(t) )  
               + 2  ~(  {\vec \pi}^{(1)}(x) \cdot {\vec \pi}^{(0)}(t)  ) ~\} .     
  \label{bg}
\end{eqnarray}
These simultaneous equations of  $  \pi^{(1)}_1 ,   \pi^{(1)}_2 ,  \pi^{(1)}_3  $ and
    $ \chi^{(1)} $  are so complicated that it is difficult to handle.
However, such complicated equations can be rewritten in a simple form
   if we represent the fluctuations $ \vec \pi^{(1)}$
   by new coordinates  $ {\pi'}^{(1)}_i$ introduced as follows.
The vector $ \vec \pi^{(1)}$ is expressed with an orthonormalized basis 
   $ ( ~{\vec e_1 }, {\vec e_2 },{\vec e_3 } ~) $,
\begin{equation}
    {\vec \pi}^{(1)} = \pi^{(1)}_1 ~ {\vec e_1} +  \pi^{(1)}_2 ~ {\vec e_2}
              + \pi^{(1)}_3 ~ {\vec e_3} .
  \label{bh}
\end{equation}
Now, introducing a new orthonormalized basis
    $ ( ~{\vec {e'_1} }, {\vec {e'_2} },{\vec {e'_3} } \equiv {\vec n} ~) $,
   one can also express $  {\vec \pi}^{(1)} $  as
\begin{equation}
    {\vec \pi}^{(1)} = {\pi '}^{(1)}_1 ~ {\vec  {e'_1} } +  {\pi '}^{(1)}_2 ~ {\vec  {e'_2} }
                          + {\pi '}^{(1)}_3 ~ {\vec  {e'_3} }.
  \label{bi}
\end{equation}
A relation between ${\vec e_i}$ and ${\vec {e'_i}}$ is
\begin{eqnarray}
  {\vec {e'_i} }~ \cdot {\vec {e_j}} &=&  
    \left(
      \begin{array}{ccc}
       \cos \theta \cos \phi  \hskip0.5cm  &  \cos \theta \sin \phi 
                                            \hskip0.5cm  &  - \sin \theta   \\
      - \sin \phi   &    \cos \phi    &   0      \\
       \sin \theta \cos \phi   &  \sin \theta \sin \phi   &  \cos \theta
     \end{array}
           \right) _{ij}                                     \nonumber              \\
  & \equiv & (M)_{i j} ~,
  \label{bj}
\end{eqnarray}
where $M$ is an orthogonal matrix.
If $ {\vec n} = {\vec  {e_3} } $, we choose  $ {\vec  {e_i} }= {\vec  {e'_i} } $ and 
    $ M_{ij} = \delta_{ij} $. 
The new coordinates $  {\pi '}^{(1)}_i $ are expressed by the orthogonal matrix  $M$,
\begin{equation}
     {\pi '}^{(1)}_i = \sum_{j=1}^{3} (M)_{i j}  ~\pi^{(1)}_j ~.
  \label{bk}
\end{equation}
In what follows, we  always write $\sum_i$ when the isospin index is summed.
The equations of motion are given in terms of the new coordinates  $ {\pi'}^{(1)}_i$,
\begin{eqnarray}
   \left[ \partial^2 + m_\pi^2 + \lambda'_1 ~ \{  \pi^{(0)} (t) \}^2 \right] 
                            {\pi '}^{(1)}_1  &=& 0 ,   \label{bla}  \\
   \left[ \partial^2 + m_\pi^2 + \lambda'_2 ~ \{  \pi^{(0)} (t) \}^2  \right] 
                           {\pi '}^{(1)}_2  &=& 0 ,     \label{blb} \\
   \left[ \partial^2 + m_\pi^2 + \lambda'_3 ~ \{  \pi^{(0)} (t) \}^2  \right] 
                                {\pi '}^{(1)}_3  &=&      \label{blc}  
               - 2 \lambda  f_\pi  ~ \pi^{(0)}(t)  \,  \chi^{(1)}  ,
\end{eqnarray}
and
\begin{equation}
  \hskip1.5cm   \left[ \partial^2 + m_\sigma^2 + \lambda ~ 
                                        \{  \pi^{(0)} (t) \}^2  \right]   {\chi}^{(1)} 
           =    -  \lambda  f_\pi   \left[  \,  \{  \pi^{(0)} (t) \}^2
                          + 2 \pi^{(0)}(t) \,  {\pi '}^{(1)}_3   \, \right]  ,  
  \label{bm}
\end{equation}
where "coupling constants" $ \lambda'_i $ are defined as
\begin{equation}
  ( \lambda'_1 , ~\lambda'_2 , ~ \lambda'_3  ) = 
                          ( \lambda , ~ \lambda , ~  3 \lambda  ).
  \label{bn}
\end{equation}
In these equations, since $  {\pi '}^{(1)}_1 $, $  {\pi '}^{(1)}_2 $ and $  {\pi '}^{(1)}_3 $
    are independent  from each other, one can analyze the equations of motion
   much easier
   than those expressed with $  {\pi }^{(1)}_i $.
The $  {\pi '}^{(1)}_3 $ is the component in the direction $\vec n$ in isospace.
It is noteworthy that, while $  {\pi '}^{(1)}_{1,2} $ are the components perpendicular
   to  $\vec n \,$,  $  {\pi '}^{(1)}_{1,2} $  ~"couple" to $  {\pi }^{(0)} (t) $ .

If $  {\vec \pi }^{(1)} $ grows very large, the approximated equations of motion
    Eqs.~(\ref{bf}) and (\ref{bg})  do not hold, because we derive these equations
   neglecting the terms which are quadratic
   and cubic in the  $  {\vec \pi }^{(1)} $ or $ \chi^{(1)} $ fields.
When $  {\vec \pi }^{(1)} $ becomes large, one should take the back reaction into
   account, i.e., the effects of the higher terms in the $  {\vec \pi }^{(1)} $ field.
In this paper, we consider the case where the amplitude $D$
   of $  {\pi }^{(0)} (t) $ is small.
The smaller the amplitude $D$ is,  the better the approximation of
   Eqs.~(\ref{bf}) and (\ref{bg})  becomes.
This will be discussed in the last section VI.
In the later sections III, IV and V, the effects of the back reaction are not considered.
%
%
%
%
%
\section{Quantum state of the DCC}
In this section, we derive the quantum state of the DCC.
The conventional picture of the DCC decay is as follows.
After quenching, the field configuration $ {\vec \pi }^{(0)} $ violating the isospin
   symmetry is realized, which is called DCC.
The field configuration then relaxes to the true vacuum $ ( ~f_\pi ,  ~{\vec 0} ~)$
    with the emission of pions.
We shall follow this conventional picture.
At first, we assume that the classical field configuration
   $\phi=( ~f_\pi ,  ~{\vec \pi }^{(0)}(t) ~) $
   violating the isospin symmetry is formed, where
   $  {\vec \pi}^{(0)}(t)  =  \pi^{(0)}(t) ~{\vec n} $
   points the direction ${\vec n} \, $,  Eq.~(\ref{beb}).
The next assumption is that the classical DCC, $\phi=( ~f_\pi ,  ~{\vec \pi }^{(0)}(t) ~) $,
   relaxes to the true vacuum  $ ( ~f_\pi ,  ~{\vec 0} ~)$,
\begin{equation}
  {\vec \pi }^{(0)}(t) \rightarrow 0 ~,  \hskip1cm ( t \rightarrow \infty ) .
  \label{ca}
\end{equation}
We study the quantum fluctuations around the field configuration
    $\phi=( ~f_\pi ,  ~{\vec \pi }^{(0)}(t) ~) $,
   that is, $ {\vec \pi }^{(0)}(t) $ is regarded as the classical back ground and we
   quantize  $ {\pi '}_i^{(1)} $. 

In the equations of motion for quantum fluctuations ${ \pi '}_i^{(1)} $ and
    $ \chi^{(1)} $, one can consider  the term ~
   $  m^2_{\pi, \sigma} + \lambda'_i ~ \{  \pi^{(0)} (t) \}^2 $ 
   as a "mass parameter" depending on the time $t$.
The violation of the isospin symmetry arises from the different behavior between 
   $  { \pi'}^{(1)}_3 $ and $  { \pi'}^{(1)}_{1,2} $ in Eqs.~(\ref{bla}) -- (\ref{bm}) .
Namely, (a) the "coupling constant" $ \lambda'_3$  is three times as much as
   $ \lambda'_{1,2}$  and
   (b) among $ ( { \pi'}^{(1)}_1, { \pi'}^{(1)}_2, { \pi'}^{(1)}_3 ) $, the component
   $ { \pi'}^{(1)}_3 $ only  couples to the sigma field $\chi^{(1)} $.
While the sigma field $\chi^{(1)} $ has no isospin, it can have a influence upon
   the isospin violation  through the interaction with $ { \pi'}^{(1)}_3 $ .
In this paper, however, we focus on only the fact (a) 
   for simplicity.
The $\chi^{(1)} $'s effect on the component  $ { \pi'}^{(1)}_3 $ is not treated, i.e., we
   set  the  right-hand side (rhs) of Eq.~(\ref{blc}) zero.
If the interaction between  $ { \pi'}^{(1)}_3 $ and  $\chi^{(1)} $ is taken into
   consideration,  the equations of motion are too complicated to handle.
The effect of (b) will be discussed in the last section VI.
Since we do not consider the contribution of  $\chi^{(1)} $ and set the rhs of
   Eq.~(\ref{blc}) zero,
   the equations of motion for $ { \pi'}^{(1)}_i $ in the momentum space become
\begin{equation}
   \left[ \frac{ \partial^2 }{ \partial t^2 } + \Omega_i ( \k, t )^2 \right]  
      {\pi '}^{(1)}_i (\k, t ) =0,   \hskip1cm  (i=1,2,3),
  \label{cb}
\end{equation}
where
\begin{equation}
    \Omega_i ( \k, t )^2  \equiv  \k^2 + m_\pi^2 +   \lambda'_i ~ \{  \pi^{(0)} (t) \}^2 .
  \label{cc}
\end{equation}
The isospin violation comes from  $ \lambda'_3 \ne \lambda'_{1,2} $ .

The explicit form of the quantum state of DCC can be derived due to neglect of the
   higher order
   terms of the quantum fluctuations and the contribution of  $\chi^{(1)} $.
Here we do not practice the self-consistent approximation, e.g., the Hartree-Fock
   approximation,  which will be a future task.
The classical field configuration $  \pi^{(0)} (t) $ is then regarded as given.
%
%
%
\subsection{Hamiltonian}
Expanding the pion fields as
\begin{equation}
   {\pi '}^{(1)}_i ( \x, t ) = \int \frac{d^3 k}{ (2 \pi)^3 } e^{i \k \x} ~{\hat Q}_i (\k, t) ,
  \label{cd}
\end{equation}
we have the Hamiltonian,
\begin{equation}
  H = \displaystyle \sum_{i=1}^3 \int d^3 k 
    \left[~
   \frac{1}{2} {\hat P}_i ( \k, t  )  {\hat P}_i ( -\k, t ) + 
   \frac{1}{2}\Omega_i ( \k, t) ^2 ~
   {\hat Q}_i (\k, t)  {\hat Q}_i ( -\k, t)        \right],
   \label{ce}
\end{equation}
from which the equations of motion Eq.~(\ref{cb}) are derived.
This Hamiltonian describes the oscillators with time dependent  $ \Omega_i ( \k, t) $.
A quantization formalism of such system is presented by Hiro-oka and Minakata
   \cite{rf:HirMinC61},   and we shall follow it.
We write as
\begin{eqnarray}
  {\hat Q}_i (\k, t) & = & Q_i (\k, t) ~{a'}_i (\k) + Q_i ( \k, t)^* ~{a'}_i^{\dagger} (-\k) , 
                                                               \nonumber  \\
  {\hat P}_i (\k, t) & = & {\dot Q}_i (\k, t) ~{a'}_i (-\k) 
                           + {\dot Q}_i (\k, t)^* ~{a'}_i^{\dagger} (\k) ,
  \label{cf}
\end{eqnarray}
where $Q_i (\k, t) $ is a complex solution of
\begin{equation}
     \left[ \frac{ \partial^2 }{ \partial t^2 } + \Omega_i ( \k, t )^2 \right] 
               Q_i ( \k, t)   =0,   \hskip1cm  (i=1,2,3).
  \label{cg}
\end{equation}
$ {a'}_i  $ and $ {a'}_i^{\dagger} $ are the time independent annihilation and 
   the creation operators, respectively.
If the Wronskian condition
\begin{equation}
  Q_j ~ \dot{Q}^*_j - \dot{Q}_j   Q_j^* =i ,
  \label{ch}
\end{equation}
is satisfied, one has consistent relations as
\begin{eqnarray}
    \left[ {a'}_i (\k) , {a'}_j^{\dagger} ( \itl ) \right]
                                 &=& \delta_{i j} \delta (\k - \itl ) ,  \nonumber  \\
   \left [  {\hat Q}_i (\k, t) , {\hat P}_j ( \itl, t) \right] 
                                &=& i ~\delta_{i j} \delta (\k - \itl ) .
  \label{ci}
\end{eqnarray}
Substitution of Eq.~(\ref{cf}) into Eq.~(\ref{ce}) yields
\begin{eqnarray}
  H &=&   \sum_i \int d^3k \frac{1}{2} \left[ A_i (\k, t) ~{a'}_i (\k) ~{a'}_i (-\k)
    + A_i^* (\k, t) ~{a'}_i ^{\dagger}(\k) ~{a'}_i^{\dagger} (-\k)  \right.  
                                                                         \nonumber  \\
 & & \hskip2cm  \left.  + B_i (\k, t)~ (~{a'}_i^{\dagger} (\k) ~{a'}_i (\k) 
                           + ~{a'}_i (\k)~{a'}_i^{\dagger} (\k)  ~)     \right] ,
  \label{cj}
\end{eqnarray}
where
\begin{eqnarray}
  A_i (\k, t) & \equiv & \dot{ Q}_i^2 + \Omega_i ( \k, t)^2 ~ Q_i^2 ,         \nonumber   \\
  B_i (\k, t) & \equiv & \dot{ Q}_i  \dot{ Q}_i^* + \Omega_i ( \k, t)^2 ~ Q_i Q_i^*  ~ > 0 .
  \label{ck}
\end{eqnarray}
%
%
%
\subsection{ The system at $ t \rightarrow \infty $ }
We are interested to express the DCC quantum state by the pion quanta defined on
   the true vacuum $ ( ~f_\pi ,  ~{\vec 0} ~)$ .
Let us consider the Hamiltonian when the classical field configuration
    $\phi=( ~f_\pi ,  ~{\vec \pi }^{(0)}(t) ~) $ becomes  $ ( ~f_\pi ,  ~{\vec 0} ~)$ 
   at $ t \rightarrow \infty $.
The assumption (\ref{ca}) gives 
\begin{equation}
  \Omega_1 ( \k, t )^2 = \Omega_2 ( \k, t )^2 = \Omega_3 ( \k, t )^2  
        = m_\pi^2 + \k^2  \equiv  \omega^2 , ~~~~~ (  t \rightarrow \infty ) .
  \label{cl}
\end{equation}
At this limit  $ t \rightarrow \infty $  with the choice of a solution
\begin{equation}
    Q_i (\k, t) = \frac{1}{\sqrt{2 \omega} } e^{ -i \omega t } , \hskip1cm ( i=1,2,3 ) ,
  \label{cm}
\end{equation}
$ A_i (\k, t ) $ becomes zero and the Hamiltonian has a diagonal form,
\begin{equation}
   H = \sum_i \int d^3k ~ \frac{ \omega }{2}   \left[
            {a'}_i^{\dagger} (\k) ~{a'}_i (\k) + ~{a'}_i (\k)~{a'}_i^{\dagger} (\k)     \right] .
  \label{cn}
\end{equation}
Substituting the solution $Q_i (\k, t ) $ into ${\pi'}^{(1)}_i ( \x, t) $ , 
   one gets at $ t \rightarrow \infty $ ;
\begin{equation}
  {\pi'}^{(1)}_i ( \x, t) = \int \frac{d^3 k}{ (2 \pi)^3 \sqrt{ 2 \omega } } \left[
       {a'}_i (\k)  e^{i ( \k \x - \omega t ) }
             +  {a'}_i ^{\dagger} (\k)  e^{ -i ( \k \x - \omega t ) } \right].
  \label{co}
\end{equation}
One can easily see that $ {a'}_i (\k) $ introduced in Eq.~(\ref{cf}) are the operators
   which diagonalize the Hamiltonian if the classical back ground field
   vanishes, $ {\vec \pi }^{(0)}(t) \rightarrow 0 $ .

Now, the charged pion's quanta  $\pi^{\pm}$ or the neutral one  $\pi^{0}$ have
   an immediate connection with the field  $ \pi_i^{(1)} $, 
   not with  $ {\pi'}_i^{(1)} $.
At the limit  $ {\vec \pi }^{(0)}(t) \rightarrow 0 ~ (  t \rightarrow \infty ) $ ,
   the  $ \pi_i^{(1)} $ obeying the equation of motion (\ref{bf}) becomes a free
   field and can be expanded as
\begin{eqnarray}
    \pi^{(1)}_i ( \x, t) &=& \int \frac{d^3 k}{ (2 \pi)^3 \sqrt{ 2 \omega } } \left[
       a_i (\k)  e^{i ( \k \x - \omega t ) } 
        +  a_i ^{\dagger} (\k)  e^{ -i ( \k \x - \omega t ) } \right] ,  \nonumber  \\
  & &    [ a_i (\k) , a_j^{\dagger} ( \itl ) ] = \delta_{i j} \delta (\k - \itl ) .
  \label{cp}
\end{eqnarray}
The Fock vacuum state $ \vert \, 0 \, \rangle $ is defined by use of $ a_i  (\k) $,
\begin{equation}
  a_i (\k) \vert \, 0 \, \rangle  = 0 ,
  \label{cq}
\end{equation}
namely,  $ \vert \, 0 \, \rangle  $ has been defined on the true vacuum configuration
    $\phi=( ~f_\pi ,  ~{\vec 0} ~) $.
A one-particle state of the charged pion $ \pi^{\pm}$ ( the neutral pion $\pi^0$ )
   with moment $\k$ is represented by 
   $  {a}_{\pm}^\dagger (\k) \vert \, 0 \, \rangle  \, 
           ( \, {a}_{3}^\dagger (\k) \vert \, 0 \, \rangle  \, )$ ,
   where
\begin{equation}
   {a}_{\pm} (\k)  \equiv  \frac{1}{\sqrt{2} } \left\{ \mp {a}_1 (\k) 
          + i ~{a}_2 (\k) \right\} .
  \label{cr}
\end{equation}

The relation between these operators  $  {a}_i (\k)  $ and $   {a'}_i (\k)   $
   introduced in Eq.~(\ref{cf}) will be necessary, for we are interested in expressing
the DCC quantum state by $  {a}_{\pm}^\dagger (\k) $ and ${a}_{3}^\dagger (\k) $.
The relation (\ref{bk}) yields
\begin{equation}
   {a'}_i (\k) = \sum_j  ( M )_{i j}~{a}_j (\k) ,
  \label{cs}
\end{equation}
and this also leads to ~$    {a'}_i (\k) \vert \, 0 \, \rangle  = 0  $ .
%
%
%
\subsection{ Derivation of quantum state }
With nonvanishing classical DCC $ {\vec \pi }^{(0)}(t) \ne 0 $ , the Hamiltonian
   (\ref{cj}) has nondiagonal terms because of
   $ A_i ( \k, t) \ne 0$ .
As stressed by Amado and Kogan  \cite{rf:AmaKog}, a squeezed state arises from the
   nondiagonal terms $ a'_i ~ a'_i ~$ and $ a^{' \dagger}_i ~a^{' \dagger}_i ~$ 
   in the Hamiltonian.
In order to find a energy eigenstate, we consider the Bogolyubov transformation;
\begin{eqnarray}
  b_i (\k,t) &=& S_i (\k,t) ~{a'}_i (\k) ~S_i^{\dagger}(\k,t)             \nonumber  \\
    &=& \cosh \theta_i (\k,t) ~{a'}_i (\k) ~
            +  e^{-i \varphi_i } \sinh \theta_i (\k,t) ~{a'}_i^{\dagger} (-\k) ,  \nonumber  \\
 b_i^{\dagger} (\k,t) &=& S_i (\k,t) ~{a'}_i^{\dagger} (\k) ~S_i^{\dagger} (\k,t)                 \nonumber  \\
   & = & \cosh \theta_i (\k,t) ~{a'}_i^{\dagger} (\k) ~+  e^{i \varphi_i } \sinh \theta_i (\k,t) ~{a'}_i (-\k) ,
  \label{ct}
\end{eqnarray}
where $ S_j (\k,t) $ is a unitary operator ;
\begin{eqnarray}
   S_j (\k,t)  & \equiv & \exp \left[ \{ \theta_j (\k,t)  ~e^{i \varphi_j  (\k,t) } \}
                 ~{a'}_j (\k)~{a'}_j (-\k) ~   \right.       \nonumber  \\
  & & \hskip1,5cm  \left.      - \{ \theta_j (\k,t)  ~e^{-i \varphi_j  (\k,t) } \}~
                   {a'}_j^{\dagger}  (\k)~{a'}_j^{\dagger} (-\k) ~ \right] .
  \label{cu}
\end{eqnarray}
If one chooses the parameters  $ \theta_j  (\k,t) $ and $ ~ \varphi_j  (\k,t) $ as
\begin{eqnarray}
   \tanh 2 \theta_j (\k,t) &=& \frac{ \vert  A_j (\k, t) \vert} {  B_j (\k, t) } , 
          \hskip1cm ( ~\theta_j (\k,t) \ge 0 ~) ,              \nonumber   \\
  A_j (\k, t) & \equiv & \vert  A_j (\k, t) \vert ~
                    \exp \left\{ i ~ \varphi_j (\k, t) \right\} ,  
 \label{cv}
\end{eqnarray}
the Hamiltonian is diagonalized \cite{rf:HirMinC61},
\begin{equation}
  H = \sum_{i=1}^3  \int d^3 \k ~ \frac{1}{2} ~\Omega_i (\k,t)  
        \left\{ ~b_i^{\dagger} (\k,t) ~b_i (\k,t)
                + b_i (\k,t) ~b_i^{\dagger} (\k,t)  \right\}.
  \label{cw}
\end{equation}
The eigenstate $ \vert  \Phi (t) \rangle $ of $H$ ,
\begin{equation}
   {b}_i (\k,t) \, \vert  \Phi (t) \rangle  = 0,     
  \label{cx}
\end{equation}
can be written as
\begin{eqnarray}
  \vert  \Phi (t) \rangle &=& \prod_{\k}   S_1(\k,t)~ S_2(\k,t)~ S_3(\k,t)~
                     \vert \,  0 \, \rangle                   \nonumber  \\
   &=&   \exp \biggl\{  \sum_{j=1}^{3}   \int d^3 \k
            \left[ \{ \theta_j (\k,t)  ~e^{i \varphi_j } \}~{a'}_j (\k)~{a'}_j (-\k) ~
                                 \right.  \biggr.                            \nonumber  \\
   & & \hskip2.8cm  \biggl. \left.        - \{ \theta_j (\k,t)  ~e^{-i \varphi_j } \}~
                           {a'}_j^{\dagger}  (\k)~{a'}_j^{\dagger}  (-\k) ~ \right] 
           \biggr\} ~ \vert \,  0 \, \rangle   .   \nonumber   \\
  \label{cy}
\end{eqnarray}
This state is known as the squeezed state where  $ \theta_j (\k,t)  $  and
   $ \varphi_j (\k,t)   $ are called the squeezing parameter and the phase parameter,
   respectively.
We regard this $  \vert  \Phi (t) \rangle  $ as the quantum state of the DCC.
The expectation value
\begin{equation}
  \langle \Phi  (t) \, \vert \, {a'}_i^{\dagger} (\k) ~{a'}_j (\k) \,
               \vert \,  \Phi (t) \rangle
  = \delta_{i j}  \, \sinh^2 \theta_i (\k,t) ,
  \label{cza}
\end{equation}
shows that the mean number of quanta $ {a'}_i^{\dagger} (\k) $ in the state
   $  \vert \,  \Phi (t) \rangle $ is given by $ \sinh^2 \theta_i (\k,t) $.
If the Wronskian condition is satisfied, one has \cite{rf:HirMinC61},
\begin{equation}
  \sinh^2 \theta_i (\k,t) = 
      \frac{  B_i (\k, t) - \Omega_i (\k, t) }{2~ \Omega_i (\k, t) }.
  \label{czb}
\end{equation}
The time development of $ \theta_j (\k,t)  $ or  $ \varphi_j  (\k,t)   $ is determined by
   the equation of motion of $ Q_j(\k,t) $ .
Since $ \Omega_1 (\k,t) = \Omega_2 (\k,t) $, we get
\begin{equation}
  \theta_1 (\k,t) = \theta_2 (\k,t) , \hskip1cm  \varphi_1 (\k, t) = \varphi_2 (\k, t) .
  \label{czc}
\end{equation}
By use of the relation (\ref{cs}), the DCC quantum state
   $  \vert \,  \Phi (t) \rangle $ is represented by $ a_{\pm}^{\dagger} (\k) $,~
   $ a_3^{\dagger}  (\k) $, polar angles $ ( \theta,  \phi ) $,
    $ \theta_j (\k,t)  $  and  $ \varphi_j  (\k,t)   $ reflecting the dynamics of the
   system.
One can easily check that 
   $   \vert \,  \Phi (t= \infty ) \rangle = \vert \, 0 \, \rangle  $.
Obtained DCC quantum state (\ref{cy}) has
   the following characteristic features:
\begin{enumerate}
  \item  The quantum state  $   \vert \,  \Phi (t ) \rangle $ contains not only the 
              component ${a'}_3^{\dagger} $ of pion quanta  in the classical DCC
              ${\vec \pi}^{(0)} $
              direction $  \vec n $ , but also  the component ${a'}_{1,2}^{\dagger} $ 
              in the perpendicular
              direction to $ \vec n $. 
  \item   The quantum state  $   \vert \,  \Phi (t ) \rangle $ is not isosinglet.
                Since $ \lambda'_3 \ne \lambda'_{1,2} $ in the equations of motion,
               it becomes  $ \theta_3 ( \k , t )  \ne \theta_{1,2} ( \k , t )   $.
  \item  The quantum state  $   \vert \,  \Phi (t ) \rangle $ contains almost no pion
             with large momentum  $\vert \k \vert \gg m_\pi $. If 
             $\vert \k \vert \gg  m_\pi $ , one has~
              $ \Omega_3 ( \k, t ) \approx \k^2 \approx  \Omega_{1,2} ( \k, t ) $ and
              $ \theta_3 ( \k , t )  \approx  \theta_{1,2} ( \k , t )   $.
              Because $ \Omega_i ( \k, t ) $ is almost independent of time, we have
              $ \vert \theta_i \, ( \k, t ) \vert  \ll 1  \, ( i=1,2,3 ) $.
\end{enumerate}
%
%
%
%
\section{ Charge fluctuation and the probability of the neutral fraction }
Since our DCC quantum state  $   \vert \,  \Phi (t ) \rangle $ is not isosinglet,
   there are $U(1)$ charge $Q$ fluctuations and the neutral fraction 
   $ f = N_{\pi^0} / ( N_{\pi^0} + N_{\pi^+} + N_{\pi^-} ) $ deviates from $1/3$.
These quantities are important when the DCC is concerned.
In this section, making use of the quantum state  $   \vert \,  \Phi (t ) \rangle $
   obtained in the previous section, we calculate the charge fluctuations 
   $ \langle \Phi (t) \vert \, Q^2 \, \vert  \,  \Phi (t ) \rangle $ and 
   the probability of the neutral fraction $ P ( f ) $ .

The isospin $ I_i \,    \, ( i=1,2,3 ) $ and the charge $Q$ is given as
\begin{equation}
    I_i  =  \int d^3 \k ~ I_i \, (\k)  , \hskip1.5cm   Q  = \int d^3 \k \,  Q ( \k ) ,  
  \label{da}
\end{equation}
where
\begin{eqnarray}
  I_i \, (\k)   &=& -i ~\epsilon_{i j l} ~a_j^{\dagger} (\k) ~a_l (\k)  ,      \nonumber   \\
  Q ( \k )  &=& I_3  ( \k )
          =    ~{a}_+^{\dagger} (\k) ~{a}_+ (\k) - ~{a}_-^{\dagger} (\k) ~{a}_- (\k)  
                                      \nonumber  \\
     && \hskip1cm   = N_{\pi^+}(\k) - N_{\pi^-}(\k)  .
  \label{db}
\end{eqnarray}
The total number operator  $N(\k)$ and the  neutral fraction $f(\k, t)$ are
\begin{eqnarray}
  N(\k) & \equiv & N_{\pi^0}(\k) +  N_{\pi^+}(\k) +N_{\pi^-}(\k) ,   \label{dc}     \\
  f (\k, t)  & \equiv &   \frac{ \langle \Phi (t) \vert \,  N_{\pi^0}(\k) \,
                                                     \vert  \,  \Phi (t ) \rangle  }
           {  \langle \Phi (t) \vert \,  N_{\pi^0}(\k) +  N_{\pi^+}(\k) +N_{\pi^-}(\k)  \,
                                                     \vert  \,  \Phi (t ) \rangle  }~,        \label{dd}
\end{eqnarray}
where $  N_{\pi^0}(\k) = {a}_3^{\dagger} (\k)  \, {a}_3 (\k) $.
For the later calculations, it is useful to define the following operators
\begin{equation}
  I'_i  \, (\k)  \equiv  -i ~\epsilon_{i j l}  \, {a'}_j^{\dagger} (\k) ~{a'}_l (\k)  ,             
  \label{de}
\end{equation}
whose relation with $  I_i  \, (\k)  $ is
\begin{equation}
   I_i  (\k) = \sum_{j=1}^3  ( M^{\rm T} )_{i j}~{I'}_j (\k).
  \label{dfa}
\end{equation}
The unit vector $ \vec n $ of the classical DCC  ${\vec \pi}^{(0)} $ points in some 
   fixed direction in isospace.
Before examining the general case of 
   ${\vec n}= ( \sin \theta \cos \phi ,  \sin \theta \sin \phi, \cos \theta ~)$,
   we consider the simple case of $ {\vec n } = ( 0,0,1 ) $.
In this case, the form of $  \vert  \,  \Phi (t ) \rangle $ becomes rather simple and
   it will help to understand how the quantum state is constructed.
%
%
%
\subsection{The case of  $ {\vec n } = ( 0,0,1 ) $ }
The classical field  ${\vec \pi}^{(0)} $ points in the direction of the neutral pion
   $\pi^0$ when  $ {\vec n } = ( 0,0,1 ) $.
We then expect that the state  $  \vert  \,  \Phi (t ) \rangle $ becomes an eigenstate
   of the charge $Q$ with an eigenvalue zero, which will be confirmed below.
Since  $ a_i = {a'}_i ~$ and  $ I_i = {I'}_i $ in the case of  $ {\vec n } = ( 0,0,1 ) $ ,
   one can use  $I'_i$ as the isospin  $I_i$.
We can easily show that 
\begin{equation}
  {I'}_3   \vert  \,  \Phi (t ) \rangle = 0 ,
  \label{dg}
\end{equation}
which means   $  \vert  \,  \Phi (t ) \rangle $ is an eigenstate  of the charge
  $Q= I_3 = {I'}_3 $ .
This fact can be seen by observing the form of  $  \vert  \,  \Phi (t ) \rangle $ which
   contains the following term,
\begin{eqnarray}
    &&  -  \{ \theta_1 (\k,t)  ~e^{-i \varphi_1 } \}~
               \left\{ ~{a}_+^{\dagger}  (\k)~{a}_-^{\dagger}  (-\k) 
                           + ~{a}_-^{\dagger}  (\k)~{a}_+^{\dagger}  (-\k)  \right\}
                              \vert \, 0 \,  \rangle            \nonumber  \\
   & &   \hskip1cm + \{ \theta_3 (\k,t)  ~e^{-i \varphi_3 } \}~~
                 {a}_3^{\dagger}  (\k)~{a}_3^{\dagger}  (-\k)   \vert \, 0 \,  \rangle ,
  \label{dh}
\end{eqnarray}
invariant under ${a}_+^{\dagger}  \leftrightarrow  {a}_-^{\dagger}  $ .
The pairs $ ( \pi^+ , \pi^- ) $ and $ ( \pi^0 , \pi^0 ) $  are contained in the state
   , but $ ( \pi^+ , \pi^0 ) $  or $ ( \pi^+ , \pi^+ ) $ is not.
The isospin of the state is
\begin{eqnarray}
  &&    \langle \, \Phi (t) \, \vert \,  {\vec I'} (\k)~\vert \,  \Phi(t) \rangle  = 0 ,   
                                   \nonumber     \\
  &&  \langle \, \Phi (t) \, \vert \,  { I'_1}^2 ~\vert \,  \Phi (t) \rangle
         =  \langle \, \Phi (t) \, \vert \,  { I'_2}^2 ~\vert \,  \Phi (t) \rangle ,
  \label{di}
\end{eqnarray}
and the neutral fraction is
\begin{equation}
          f (\k, t) = \frac{ \sinh^2 \theta_3 (\k,t) } 
             { \sum_{i=1}^3 \sinh^2 \theta_i \, (\k,t) } .
  \label{dj}
\end{equation}
These calculations are useful in the general case of 
    ${\vec n}= ( \sin \theta \cos \phi ,  \sin \theta \sin \phi, \cos \theta ~)$.
%
%
%
\subsection{ The case of 
        ${\vec n}= ( \sin \theta \cos \phi ,  \sin \theta \sin \phi, \cos \theta ~)$  }
The expectation values of $ {\vec I} (\k) $ and $ Q(\k)=I_3 (\k) $ are zero,
\begin{equation}
    \langle \, \Phi(t) \, \vert \,  {\vec I} (\k)~\vert \,  \Phi (t) \rangle  = 0,
       \hskip1cm
    \langle \, \Phi (t) \, \vert \,  Q (\k)~\vert \,  \Phi (t) \rangle  = 0 .
  \label{dk}
\end{equation}
However, $ \vert \,  \Phi (t) \rangle $ is not an eigenstate of the charge $Q$
   in general.
With the help of the relation, 
   $  Q  = \sum_{j=1}^3  ( M^{\rm T} )_{3 j}~{I'}_j  $ , 
   the charge fluctuation can be calculated as
\begin{eqnarray}
   & & \hspace{-0.3cm}   \langle \, \Phi(t) \,  \vert \,   Q^2 \, 
            \vert   \,  \Phi (t) \rangle                  \nonumber  \\
   &=&  \sin^2 \theta  \times  \int d^3 \k  
            \biggl[    \cosh^2 \theta_2 (\k,t) ~ \sinh^2 \theta_3 (\k,t) 
                      +   \cosh^2 \theta_3 (\k,t) ~ \sinh^2 \theta_2 (\k,t)    \biggr.    
                                    \nonumber  \\
   & &   \biggl.  -  \cosh \theta_2 (\k,t) ~ \sinh \theta_2 (\k,t) 
                        \cosh \theta_3 (\k,t) ~ \sinh \theta_3 (\k,t) 
                \times 2 \cos \{ \varphi_3 (\k,t)  - \varphi_2 (\k,t) \}
                          \biggr].   \nonumber  \\
  \label{dl}
\end{eqnarray}
The relation between the fluctuation of the charge and that of the isospin is
\begin{equation}
   \langle \, \Phi (t) \, \vert \,  Q^2 ~\vert \,  \Phi (t) \rangle
   =  \frac{\sin^2  \theta  }{2}  \langle \, \Phi (t) \, \vert \,   {\vec I}^2 ~
           \vert \,  \Phi (t)  \rangle ,
  \label{dm}
\end{equation}
where Eqs.~(\ref{dfa}) and (\ref{di}) have been used. 

The neutral fraction is calculated by use of Eq.~(\ref{cs}),
\begin{equation}
   f(\k, t) = \alpha_{\parallel} (\k, t) \cos^2 \theta
               + \alpha_{\perp} (\k, t)  \sin^2 \theta  ,
  \label{dn}
\end{equation}
where we have defined
\begin{eqnarray}
   \alpha_{\parallel} (\k, t) & \equiv &
        \frac{  \langle \Phi  (t) \, \vert \, {a'}_3^{\dagger} (\k) ~{a'}_3 (\k) \,
                                 \vert \,  \Phi (t) \rangle }
               { \sum_{i=1}^3   \langle \Phi  (t) \, \vert \, {a'}_i^{\dagger} (\k) ~{a'}_i (\k) \,
                               \vert \,  \Phi (t) \rangle  }
       = \frac{    \sinh^2 \theta_3 (\k,t)  }  { \sum_{i=1}^3 \sinh^2 \theta_i (\k,t) } ~,
                                                                           \label{do}   \\
   \alpha_{\perp} (\k, t) & \equiv &
        \frac{  \langle \Phi  (t) \, \vert \, {a'}_1^{\dagger} (\k) ~{a'}_1 (\k) \,
                                 \vert \,  \Phi (t) \rangle }
               { \sum_{i=1}^3   \langle \Phi  (t) \, \vert \, {a'}_i^{\dagger} (\k) ~{a'}_i (\k) \,
                               \vert \,  \Phi (t) \rangle  }
       = \frac{    \sinh^2 \theta_1 (\k,t)  }  { \sum_{i=1}^3 \sinh^2 \theta_i (\k,t) } ~.
                                                                           \label{dp}  
\end{eqnarray}
The $  \alpha_{\parallel} (\k, t)  $  (  $  \alpha_{\perp} (\k, t)  $  ) represents the
   contribution of the pion component in the direction $\vec n$
   (  in the perpendicular direction to $\vec n$ ).
If the state $   \vert \,  \Phi (t) \rangle $ contained no pion component 
    in the perpendicular direction to $\vec n$ , it would become 
   $ \alpha_{\parallel} (\k, t) =1 ,~ \alpha_{\perp} (\k, t) =0 $, and 
   $ f  (\k, t) = \cos^2 \theta $.
When  $ \sinh^2 \theta_1 \le  \sinh^2 \theta_3 $
   (  $ \alpha_{\perp} \le 1/3 \le  \alpha_{\parallel} \, $ ),
   the neutral fraction is within the range 
   $ \alpha_{\perp} \le f \le  \alpha_{\parallel} \, $.  
We expect all directions $ {\vec n} = ( x, y, z ) $ in isospace are equally likely, then
   the  probability of the neutral fraction becomes \cite{rf:AmaKog} 
\begin{eqnarray}
  P(f(\k, t) ) &=& N \int dx dy dz ~ \delta ( 1- x^2-y^2-z^2 )       \nonumber  \\
  && \hskip0.8cm  \times 
     \delta \left( f - \frac{ z^2}{ x^2+y^2+z^2 } ~  \alpha_{\parallel}
                             - \frac{ (y^2 +z^2) }{ x^2+y^2+z^2 } ~  \alpha_{\perp}   \right)    
                                                                     \nonumber  \\
  & = & \frac{1}{ 2 \sqrt{  \alpha_{\parallel} (\k, t) -  \alpha_{\perp} (\k, t)  } } 
           ~\frac{1}{ \sqrt{ f (\k, t) -  \alpha_{\perp} (\k, t) } } .
  \label{dq}
\end{eqnarray}
The $P(f)$ has a maximum at $f=\alpha_{\perp}$ and a minimum
   at  $f=\alpha_{\parallel}$.
The charge fluctuation 
    $ \langle \, \Phi (t) \, \vert \,  Q^2 ~\vert \,  \Phi (t) \rangle $
   and  $P (f(\k, t) ) $  
  are determined by
   the squeezing parameter $ \theta_i \, (\k,t) $ and the phase parameter
   $ \varphi_i (\k,t) $ which are known from the equations of motion (\ref{cg}) .

Before investigating the dynamics for  $ \theta_3 \, (\k,t) $ and 
    $ \theta_1 \, (\k,t) =\theta_2 \, (\k,t) $ , we shall consider two extreme cases.
The first case is that $ \theta_3 = \theta_1 ( =\theta_2 ) $ and
    $ \varphi_3 = \varphi_1 ( =\varphi_2 ) $.
In this case, one obtains
\begin{eqnarray}
   &&  \langle \, \Phi (t) \, \vert \,  Q^2 \, \vert \,  \Phi (t) \rangle = 0 ~,    \\
   &&  f( \k, t ) = \frac{1}{3}~ ,  \hskip0.7cm  
            (  ~ \alpha_{\parallel} =  \alpha_{\perp} = \frac{1}{3} ~) ,  
  \label{dr}
\end{eqnarray}
which are natural results in the isospin symmetric case.
Because of $ \Omega_3 \ne \Omega_{1,2} $  $(  \lambda'_3 \ne \lambda'_{1,2} )$
   in the equations of motion, $\theta_3$ is not equal to  $\theta_{1,2}$ in general.
However, such extreme case will be realized approximately if 
   $ t \rightarrow \infty $
   $ (  \Omega_3 \approx \sqrt{ \k^2+m_\pi^2 } \approx \Omega_{1,2} ) $ or
   $ \vert \k \vert \gg m_\pi $ 
   $ (  \Omega_3 \approx \vert \k \vert   \approx \Omega_{1,2} ) $.
The second case is that 
   $  \sinh^2 \theta_3 (\k,t)   \gg   \sinh^2 \theta_{1,2} (\k,t)  ~ $ and
   $   \sinh^2 \theta_3 (\k,t)   \gg   1 $.
In this case, the charge fluctuation is small,
\begin{equation}
     \langle \, \Phi (t) \, \vert \,  { Q}^2 ~\vert \,  \Phi (t) \rangle   \ll 
          \langle \, \Phi (t) \, \vert \,  N ~\vert \,  \Phi (t) \rangle^2  .
  \label{ds}
\end{equation}
Moreover, we have
\begin{equation}
     f( \k, t ) \approx \cos^2 \theta ~ ,  \hskip0.7cm  
            (  ~ \alpha_{\parallel} \approx 1 , ~  \alpha_{\perp} \approx 0 ~) ,  
  \label{dt}
\end{equation}
by which
\begin{equation}
  P(f) \approx \frac{1}{ 2 \sqrt{ f } } .
  \label{du}
\end{equation}
Therefore, if the relation 
     $  \sinh^2 \theta_3 (\k,t)   \gg   \sinh^2 \theta_{1,2} (\k,t)  ~ $is realized,
   the probability $P(f)$ becomes the well known form.
In the next section, we consider the dynamics for $\theta_i (\k,t)  $.
%
%
%
%
%
\section{ Parametric resonance for low momentum pions }
If the $P(f)$ shows some deviation from $\delta \, ( f - 1/3 \,) $ in the experiments,
   it suggests that the DCC is formed.
It is desirable that the parameters take the value
   $  ~ \alpha_{\parallel} \approx 1 $ and $~  \alpha_{\perp} \approx 0 ~$
   in Eq.~(\ref{dq}) 
   so as to distinguish the probability distribution $P(f)$ clearly from 
   $\delta \, ( f - 1/3 \,) $.   
As mentioned in the previous section, it becomes  $ P(f) \approx 1/ 2 \sqrt{f} $
   if the great difference 
    $  \sinh^2 \theta_3 (\k,t)   \gg   \sinh^2 \theta_{1,2} (\k,t) $ is realized by the
   dynamics.
The time development of  $  \sinh^2 \theta_3 (\k,t) $ differs from that of
   $   \sinh^2 \theta_{1,2} (\k,t) $ , because the coupling constant $ \lambda'_3  $ 
   is not equal to $  \lambda'_{1,2} $ in the equations of motion (\ref{cg}) , i.e.,
    $ \lambda'_3 = 3 \lambda'_{1,2} $ $( \Omega_3 \ne \Omega_{1,2} )$.
Is it possible that such a small difference in coupling constant gives rise to a great
   difference between  $  \sinh^2 \theta_3 (\k,t) $ and 
   $   \sinh^2 \theta_{1,2} (\k,t) $ ?
For the low momentum pion modes, it is indeed possible by the parametric resonance
   mechanism.
In this section, we show that  the parametric resonance works for low momentum
   pions and examine its effect on $P(f)$ and the fluctuation of the charge
   $   \langle \, \Phi (t) \, \vert \,  { Q}^2 ~\vert \,  \Phi (t) \rangle $.

We have assumed in Sec.~III that the classical DCC $\phi$ relaxes to the true vacuum,
   $ {\vec \pi }^{(0)}(t) \rightarrow 0 ~   (~ t \rightarrow \infty ~) $. 
At short intervals of time, however, we can approximate that $ {\pi}^{(0)} (t) $
   oscillates with a constant amplitude $D$,
\begin{equation}
   \pi^{(0)}(t)  \approx D \cos ( m_\pi ~ t ) ,
  \label{ea}
\end{equation}
then
\begin{equation}
    \Omega_i ( \k, t )^2 
     \approx  \k^2 + m_\pi^2 +   \frac{ \lambda'_i }{2} D^2 
                     +   \frac{ \lambda'_i }{2} D^2   \cos  ( 2~ m_\pi ~ t ) .
  \label{eb}
\end{equation}
Let us consider the small amplitude $D$ such that 
\begin{equation}
    m_\pi^2 \gg  \frac{\lambda'_i }{2} \, D^2  . 
  \label{ec}
\end{equation}
We  neglect the third term in Eq.~(\ref{eb}) , but not the fourth term which produces
   the parametric resonance.
Furthermore, we restrict the pion modes with  $ 0 < \vert \k \vert \ll m_\pi $ .
In what follows, $\k$ represents the low momentum  $ 0 < \vert \k \vert \ll m_\pi $ 
   in this section.
With the approximation, the equation of motion  (\ref{cg}) then becomes
\begin{eqnarray}
 0 & = &   \left[ \frac{ \partial^2 }{ \partial t^2 } +  m_\pi^2 +  
      \frac{ \lambda'_i }{2} D^2  \cos ( 2 \, m_\pi \, t )     \right]   Q_i \, ( \k , t) 
                         \nonumber  \\
  & \equiv &  \left[ \frac{ \partial^2 }{ \partial t^2 } +  m_\pi^2  \biggl\{ 
         1+ h_i   \cos ( 2 \, m_\pi \, t )   \biggr\}  \right]   Q_i ( \k , t) ,   
  \label{ed}
\end{eqnarray}
where
\begin{equation}
  h_i \equiv \frac{1}{m_\pi^2} \frac{  \lambda'_i ~ D^2 }{2} \ll 1 .
  \label{ef}
\end{equation}
Eq.~(\ref{ed}) is called the  Mathieu equation \cite{rf:LanLif}, whose unstable
   solution increases rapidly with time.
Although the initial condition must be given to obtain a solution  $ Q_i ( \k , t ) $ ,
   we do not know at the moment what the initial condition of the quantum
   fluctuation   $ {\hat Q}_i ( \k , t ) $ should be.
Here, since we are interested in whether 
    $  \sinh^2 \theta_3 (\k,t)   \gg   \sinh^2 \theta_{1,2} (\k,t) $ is realized by the
   dynamics or not, we choose the initial condition of  $ Q_i ( \k , t ) $ so as to hold~
   $  \sinh^2 \theta_i (\k,t )   = 0 ~  ( ~i = 1,2,3 ~) $ for $t=0$.
A solution of Eq.~(\ref{ed}) up to $ {\cal O }( ( h_i )^0  \,) $ is \cite{rf:LanLif}
\begin{eqnarray}
   Q_i ( \k , t) & = & \frac{1}{ \sqrt{ 2 m_\pi } } \left[
           \exp \left( \frac{ h_i ~m_\pi }{4}~ t \right) 
              \left\{  \cos \left( m_\pi t + \frac{\pi}{4} \right) + {\cal O} ( h_i  \,)  \right\} 
                         \right.       \nonumber  \\
   && \hskip1.3cm  \left.   - i  \exp \left( - \frac{ h_i ~m_\pi }{4}~ t \right)
        \left\{  \sin \left( m_\pi t + \frac{\pi}{4} \right)
                     +  {\cal O} ( h_i  \,)  \right\}   \right] ,
  \label{eg}
\end{eqnarray}
which satisfies the Wronskian condition up to $ {\cal O} ( (h_i )^0 \,) $.
The parameters defined in Eq.~(\ref{ck})  are
\begin{eqnarray}
   A_i (\k, t) &=&  m_\pi ~\sinh \left( \frac{ h_i ~m_\pi }{2}~ t \right) 
                    + {\cal O} (h_i)  ,     \nonumber  \\
   B_i (\k, t) &=&  m_\pi ~\cosh \left( \frac{ h_i ~m_\pi }{2}~ t \right)
                  + {\cal O} (h_i)    .
  \label{eh}
\end{eqnarray}
The squeezing parameter and the phase parameter are then
   obtained by Eq.~(\ref{cv}) ,
\begin{eqnarray}
   && \sinh^2  \theta_i \, (\k,t) = \sinh^2 \left( \frac{ h_i ~m_\pi }{4}~ t \right)
                  +  {\cal O}  (h_i \,) ,               \\
   && \varphi_1 = \varphi_2 = \varphi_3 = 0  +  {\cal O}  (h_i \,) .
  \label{ei}
\end{eqnarray}
Our solution (\ref{eg}) satisfies the required condition,
    $  \sinh^2 \theta_i (\k,t=0 )   = 0 $  up to $ {\cal O} ( ( h_i )^0  \,) $.
Using the obtained squeezing parameter  $  \theta_i (\k ,t) $ , we can easily
   check that ~$  \sinh^2 \theta_3 (\k,t)   \gg   \sinh^2 \theta_{1,2} (\k,t) $ 
   as follows,
\begin{equation}
   \frac{ \sinh^2  \theta_1 (\k ,t) }{\sinh^2  \theta_3 (\k ,t) }
   =  \frac{ \sinh^2   \left(  \lambda'_1  \, \frac{ D^2 t }{ 8 \, m_\pi } \right)  }
               { \sinh^2   \left(  \lambda'_3  \, \frac{ D^2 t }{ 8 \, m_\pi } \right)   }
             + {\cal O} (h_i \,)                           
   =  \frac{1} { \left[ 3+ 4 \sinh^2  \left( \frac{  \lambda ~ D^2 }{8~ m_\pi} ~ t  \right) 
              \right]^2  }  +  {\cal O} (h_i \,)  .
  \label{ej}
\end{equation}
Even when $t \approx 0$, the ratio is about $0.11$ and it decreases rapidly with time.
Thus, the small difference in coupling constant  $ \lambda'_3 = 3 \lambda'_{1,2} $ 
   produces the great difference 
   $  \sinh^2 \theta_3 (\k,t)   \gg   \sinh^2 \theta_{1,2} (\k,t) $ 
   through the parametric resonance.

By use of the solution  (\ref{eg}) which reflects the  parametric resonance,
   we shall calculate the fluctuation of the charge and the probability of the
   neutral fraction for the low momentum pion with $ 0 < \vert \k \vert \ll m_\pi $.
Note that Eq.~(\ref{eg}) is the solution of the equation of motion
   neglecting the higher order terms of the fluctuations ( back reaction terms ).
We will discuss a range of applicability which this solution has,  in Sec.~VI.
Bearing this in mind, we first calculate the charge fluctuation (\ref{dl}) of the
   low momentum pion with  $ 0 < \vert \k \vert \ll m_\pi $,
\begin{eqnarray}
  \langle \, \Phi (t) \, \vert \,  { Q}^2 \, \vert \,  \Phi (t) \rangle   
   &=&  \sin^2  \theta  \biggl[  \cosh  \left( \frac{  ~\lambda ~ D^2 }{8~ m_\pi} 
                         ~ t  \right) 
                       ~ \sinh 3  \left( \frac{  ~\lambda ~ D^2 }{8~ m_\pi} ~ t  \right) 
                            \biggr.                     \nonumber \\
  & &  \hskip1cm  \biggl.    -  \cosh 3  \left( 
                   \frac{  ~\lambda ~ D^2 }{8~ m_\pi} ~ t  \right) 
               ~ \sinh  \left( \frac{  ~\lambda ~ D^2 }{8~ m_\pi} ~ t  \right)     \biggr]^2 
                   + {\cal O} (h_i \,) .  
  \label{ek}
\end{eqnarray}
The total number of pion with low momentum  $ 0 < \vert \k \vert \ll m_\pi $ is
\begin{equation}
    \langle \Phi ( t ) \vert \,  N(\k  ) \, \vert \, \Phi ( t ) \rangle 
       \approx  \sinh^2   3 \left( \frac{   ~\lambda ~ D^2 }{8~ m_\pi} ~ t  \right) 
        + 2 \sinh^2  \left( \frac{  ~\lambda ~ D^2 }{8~ m_\pi} ~ t  \right)
          +  {\cal O} ( h_i \,) .
  \label{el}
\end{equation}
Now, with the help of an inequality,~
   $   2 \left[  \cosh x ~ \sinh 3 x  -  \cosh 3 x ~ \sinh x     \right]^2   
         <   [ ~2 \sinh^2 x +  \sinh^2 3 x  ~]   , ( x  > 0) $,  we get
\begin{equation}
    \langle \Phi ( t ) \vert \,  Q^2  \, \vert \, \Phi ( t ) \rangle \,
     \le   \,   \frac{ \sin^2 \theta }{2}   \,  \langle \Phi( t ) \vert \,  N(\k  ) \,
            \vert \, \Phi ( t ) \rangle .
  \label{em}
\end{equation}
Therefore, the  fractional root mean square charge fluctuation of the low momentum
   pion is of order 
  $ \sin \theta / \sqrt{ 2  \langle \Phi ( t ) \vert \,  N(\k  ) \, \vert \, \Phi ( t )
   \rangle } $ 
   and is negligible when the total pion number is large.
Next, we focus on $ P(f (\k ,t ) ) $.
Because of  $  \sinh^2 \theta_3 (\k,t)   \gg   \sinh^2 \theta_{1,2} (\k,t) $ ,
   one has 
\begin{eqnarray}
  \alpha_{\perp} (\k , t)  & \approx &  \frac{ \sinh^2 \theta_1 (\k,t) }
           { \sinh^2  \theta_3 (\k,t) }
      =   \frac{1} { \left[ 3+ 4 \sinh^2  \left( \frac{  \lambda ~ D^2 }{8~ m_\pi} ~ t 
                  \right)     \right]^2  } + {\cal O} (h_i \,) ,              \nonumber  \\
     \alpha_{\parallel} (\k , t)  & \approx &  1 - \frac{  2 \sinh^2 \theta_1 (\k,t) } 
              { \sinh^2  \theta_3 (\k,t) }
    =   1 - \frac{2} { \left[ 3+ 4 \sinh^2  \left( \frac{  \lambda ~ D^2 }{8~ m_\pi} ~ t 
                    \right)     \right]^2  } + {\cal O}  (h_i \,) .
  \label{en}
\end{eqnarray}
Even when $t \approx 0$, these quantities are $  \alpha_{\perp} \approx 0.11 $,
   $  \alpha_{\parallel} \approx 0.78 $ and they rapidly converge with time as
   $  \alpha_{\perp} \rightarrow 0 $~, $  \alpha_{\parallel} \rightarrow 1 $~.
Therefore, we have
\begin{equation}
       P(f  ) \approx  \frac{1}{2 \sqrt{f} } ,
  \label{eo}
\end{equation}
for the low momentum pion.

Thus, the parametric resonance mechanism works for the low momentum pion
   with $ 0 < \vert \k \vert \ll m_\pi $ with the result that the 
   fractional root mean square charge fluctuation  $ \sqrt{ < Q^2 > } / < N > $ is
   negligible when $ < N > $ is large and that the probability of the neutral fraction
   becomes
\begin{equation}
      P(f ) =   \frac{1}{ 2 \sqrt{  \alpha_{\parallel} -  \alpha_{\perp}  }  } 
                  ~\frac{1}{ \sqrt{ f -  \alpha_{\perp} }  } ~
             \approx  ~ \frac{1}{2 \sqrt{f} }~ .
  \label{ep}
\end{equation}
%
%
%
%
%
\section{Conclusion and discussion}
We studied the quantum aspect of the disoriented chiral condensate ( DCC ) by use
   of the $O(4)$ linear sigma model.
We derived a quantum state of the DCC dynamically, considering small quantum
   fluctuations around a classical DCC on the assumption that the classical chiral
   condensate is in a direction $\vec n$ in isospace.
The properties of the obtained nonisosinglet quantum state, the charge fluctuation
   and  the probability of the neutral fraction are investigated.

The obtained DCC quantum state $   \vert \, \Phi ( t ) \rangle $
   has the following characteristic features;
   (i) it has the form of the squeezed state,
  (ii) the state contains not only the component of pion quanta in the direction
      $  \vec n $ but
      also  the component in the perpendicular direction to $  \vec n $
   and (iii) the low momentum pions  with $  \vert \k \vert < m_\pi $ in the state
       violate the  isospin symmetry.   
With the quantum state $   \vert \, \Phi ( t ) \rangle $, we find the probability of
   the neutral fraction $  P(f(\k, t) ) $ becomes
\begin{equation}
  P(f(\k, t) )
   =  \frac{1}{ 2 \sqrt{  \alpha_{\parallel} (\k, t) -  \alpha_{\perp} (\k, t)  } } 
           ~\frac{1}{ \sqrt{ f (\k, t) -  \alpha_{\perp} (\k, t) } } ,
  \label{fa}
\end{equation}
where $  \alpha_{\parallel} (\k, t) $ and $ \alpha_{\perp} (\k, t) $ are determined
   by the dynamics.
For  the low momentum pions  with $  0 < \vert \k \vert \ll m_\pi $ , the parametric
   resonance mechanism works with the result that (a) the charge fluctuation
   of the pion quanta is negligible if the total pion number is large, and
   (b) it becomes  $ \alpha_{\perp} (\k, t) \approx 0$ , 
    $  \alpha_{\parallel} (\k, t) \approx 1 $ and 
\begin{equation}
        P(f(\k, t) )  \approx   \frac{1}{ 2 \sqrt{ f  } }.
  \label{fb}
\end{equation}
Now, the low and high momentum approximations for the probability of the
   neutral fraction are
\begin{eqnarray}
    P(f(\k, t) ) = 
     \left\{  \begin{array}{ll}
                 \frac{1}{ 2 \sqrt{ f  } }   \, ,                                            &
                                               \;  \mbox{  $  \vert  \k \vert \ll m_\pi  $ }     \\
                  \frac{1}{ 2 \sqrt{ \alpha_{\parallel} - \alpha_{\perp} } } 
                       ~\frac{1}{ \sqrt{ f  - \alpha_{\perp}  }  }   \, ,  &
                                         \;    \mbox{   $  \vert \k \vert  \sim  {\cal O} ( m_\pi )  $    }          \\
                 \delta (f - \frac{1}{3} )    \, ,                                        &
                                                \;  \mbox{  $  \vert \k \vert \gg  m_\pi  $ }  
                 \end{array}
     \right.
 \label{fc}
\end{eqnarray}
The probability $ P(f(\k, t) ) $ has an unfamiliar form 
   $ 1/ 2 \sqrt{ \alpha_{\parallel} - \alpha_{\perp} } \sqrt{ f  - \alpha_{\perp}  } $
   because the component of the pion quanta in the perpendicular direction to
   $\vec n$ contributes, $ \alpha_{\perp}  \ne 0$.
For  the low momentum pions  with $  0 < \vert \k \vert \ll m_\pi $ ,  $ P(f(\k, t) ) $ 
   takes the well known form $1/2 \sqrt{f} $ approximately by the parametric
   resonance mechanism.
In order to observe the DCC, it is better to measure the low momentum pions.

Because we neglect the sigma field ${\chi}^{(1)}$ and the effect of the quantum back
   reaction, the explicit form of the DCC quantum state  $   \vert \, \Phi ( t ) \rangle $
   can be obtained.
Henceforth, we discuss these effects which are disregarded in our calculations.
In Sec.~III, the sigma  ${\chi}^{(1)}$ in the equation of motion  (\ref{blc}) 
   has been neglected.
Here, let us consider the  equation of motion with  ${\chi}^{(1)}$ qualitatively in order
   to estimate the effect of  ${\chi}^{(1)}$ on the time development 
   of  ${\pi '}^{(1)}_3  ( ~Q_3 ( \k,t )  ~) $ .
Without neglecting the sigma  ${\chi}^{(1)}$ ,  the equation of motion (\ref{ed}) 
   in terms of
   $ ~Q_3 ( \k,t )  ~$ is replaced by
\begin{equation}
   \left[ \frac{ \partial^2 }{ \partial t^2 } +  m_\pi^2  \biggl\{ 
         1+ h_3   \cos\, ( 2 m_\pi ~ t )   \biggr\}  \right]   Q_3 ( \k, t)  
     =  -2 \lambda f_\pi \pi^{(0)} (t)  \,  \chi^{(1)} (\k, t) ,
  \label{fd}
\end{equation}
where  $ 0 < \vert \k \vert \ll m_\pi $ .
A general solution of this inhomogeneous equation is given by the sum of
   a homogeneous solution and a particular solution.
Since the homogeneous solution  (\ref{eg}) increases exponentially
   with time, a solution of  Eq.~(\ref{fd}) should increase rapidly with time.
Therefore,  for the low momentum mode,
   the instability of $ Q_3 ( \k,t )  $ will not be affected by including the
   sigma  ${\chi}^{(1)}$.
We will also comment on the sigma's quanta.
When the quantum state of the sigma is considered in the DCC quantum state,
   it is necessary to take into account the decay $ \sigma \rightarrow \pi \pi $
   which contributes to $ < N_\pi > $ and $ P ( f )  $ .

Next, we discuss the effect of the quantum back reaction.
If the quantum fluctuation grows to be large, one can not neglect the higher order
   terms of those fluctuation which will affect on the chiral condensate's time
   development or the expectation values of physical quantity.
In Sec.~III, calculations including the quantum back reaction were not carried out
   but we have assumed that the classical chiral condensate damps,
   $ { \vec  \pi}^{(0)}(t)   \rightarrow 0 \, ( t \rightarrow \infty ) $.
This assumption looks reasonable if the references Ref.\cite{rf:TsuKoiIke} and
   Ref.\cite{rf:HirMinC64} are considered, in which the calculations with the
   quantum back reaction are carried out.
In these references, the chiral condensate and the quantum fluctuation are treated
   in the self-consistent way, and it is shown numerically that the chiral condensate
   damps. 

Finally, we concern about the time $\tau$  the quantum back reaction takes effect.
Eq.~(\ref{eg}) is a solution of the approximated
   equation of motion neglecting the higher order terms of the fluctuation.
If  $ Q_3 ( \k,t )  $ grows to be large, that approximated equation of motion does not
   hold because of the back reaction.
The solution (\ref{eg}) behaves exponentially, and when the
   time $t$ exceeds $\tau$ such that $  ( h_i ~m_\pi / 4  \,) \, \tau  \sim 1 $,
    $ Q_3 ( \k,t )  $ grows rapidly.
The time $\tau$ is given by use of $ h_3 = 3 \lambda D^2 / 2 \, m_\pi^2 $ ,
\begin{equation}
  \tau \sim   \frac{8 \,  m_\pi} { 3 \lambda \,  D^2 }  ,
  \label{fe}
\end{equation}
until $t=\tau$, ~$ Q_3 ( \k,t )  $ is not so large.
Therefore, the quantum back reaction would not significantly affect if $t < \tau$.
As done in Sec.~V,  when we consider a small amplitude $D$ of  ${ \pi}^{(0)} (t) $ ,
   the time $\tau$ becomes enough long.
However, in order to take into account the  quantum back reaction rigorously, the
   formalism including the  quantum back reaction fully is needed.
In such a formalism, the decay time of the DCC might be obtained.
%
%
%
%
\section*{Acknowledgments}

The author would like to thank H. Hiro-oka,  M. Ishihara and H.Minakata for
   useful discussions.
%
%
%
%
%
%
%
%
%
%

%
%
%
\end{document}